\documentclass{ws-procs10x7}
\usepackage{amsmath}
\usepackage{amssymb}
\usepackage{cite}
\begin{document}

\title{Massive Elementary Particles and Black Holes in Resummed Quantum Gravity}

\author{B.F.L. Ward}

\address{Department of Physics, Baylor University, Waco, TX, USA}




\twocolumn[\maketitle\abstract{
We use exact results in a new approach to quantum gravity to 
show that the classical conclusion that a massive elementary point
particle is a black hole is obviated by quantum loop effects.
Further phenomenological implications are discussed.}]

\section{Introduction}

Albert Einstein showed that Newton's law, one of the most basic laws in
physics, is a special case of the solutions of the
classical field equations of his general theory of relativity. Specifically,
$g_{00}=1+2\Phi_N \Rightarrow \nabla^2\Phi_N=4\pi G_N\rho$
from
$R^{\alpha\gamma}-\frac{1}{2}g^{\alpha\gamma}R  =-8\pi G_N T^{\alpha\gamma}$, etc., where he have now introduced the familiar metric of space-time
$g_{\mu\nu}$, the Newtonian potential $\Phi_N$, Newton's constant
$G_N$, the mass density $\rho$, the contracted Riemann 
tensor $R^{\alpha\gamma}$,
and the appropriate energy momentum tensor $T^{\alpha\gamma}$.
There have been several successful tests of Einstein's theory in classical physics~\cite{mtw,sw1,abs}.\par

Heisenberg and Schroedinger, following Bohr, formulated a quantum mechanics
that has explained, in 
the Standard Model(SM)~\cite{sm}, 
all established experimentally accessible 
quantum phenomena except the quantum treatment of Newton's law. Indeed,
even with tremendous progress in quantum field theory,
superstrings~\cite{gsw,jp}, loop quantum gravity~\cite{lqg}, etc.,
no satisfactory treatment of the
quantum mechanics of Newton's law is known to be correct 
phenomenologically.
Here,
we apply a new approach~\cite{bw1} to quantum gravitational phenomena, 
building on previous work by Feynman~\cite{f1,f2} to get a minimal
union of Bohr's and Einstein's ideas.

There are four approaches~\cite{wein1} to the attendant bad UV behavior 
of quantum gravity (QG):
extended theories of gravitation such as supersymmetric theories - superstrings
and loop quantum gravity;
resummation, a new version of which we discuss presently;
composite gravitons; and,
asymptotic safety -- fixed point theory, recently pursued with
success in Refs.~\cite{laut,reuter2}.
Our approach allows us to make contact
with both the extended theory approach and the asymptotic safety
approach.

Our new approach 
, resummed quantum gravity, is based on well-tested YFS~\cite{yfs,yfs1}
methods. We first review
Feynman's formulation of Einstein's theory in Sect. 2. We present
resummed QG in Sect. 3. In Sect. 4 we discuss Newton's law.
In Sect. 5 we discuss the black hole physics, some of which is
related to Hawking
radiation~\cite{hawk}.

\section{Review of Feynman's Formulation of Einstein's Theory}

For the known world, we have the generally covariant Lagrangian
\begin{equation}
{\cal L}(x) = -\frac{1}{2\kappa^2}\sqrt{-g} R
            + \sqrt{-g} L^{\cal G}_{SM}(x)
\label{lgwrld}
\end{equation}
where $R$ is the curvature scalar, 
$-g=-{\text det} g_{\mu\nu} $, 
$\kappa=\sqrt{8\pi G_N}\equiv 
\sqrt{8\pi/M_{Pl}^2}$, 
where {$G_N$} is Newton's constant, and the
SM Lagrangian density is $ L^{\cal G}_{SM}(x)$.
One gets $L^{\cal G}_{SM}(x)$ from the usual SM Lagrangian density
by standard methods that are presented in Refs.~\cite{bw1}.

In the SM there are many massive point particles.
Are they black holes in our new approach to quantum gravity?
To study this question, we follow Feynman, treat spin as
an inessential complication~\cite{mlg}, and 
replace $L^{\cal G}_{SM}(x)$ in (\ref{lgwrld}) with
the simplest case for our question, that of a free scalar
field, a free physical Higgs field, $\varphi(x)$, with a rest mass believed to be less than $400$ GeV and known to be greater than $114.4$ GeV with a
95\% CL~\cite{lewwg}. We are then led to consider the representative 
model~\cite{f1,f2} {\small
\begin{equation}
\begin{split}
{\cal L}(x) &= -\frac{\sqrt{-g}}{2\kappa^2} R
            + \frac{\sqrt{-g}}{2}\left(g^{\mu\nu}\partial_\mu\varphi\partial_\nu\varphi - m_o^2\varphi^2\right)\\
            &= \frac{1}{2}{\big\{} h^{\mu\nu,\lambda}\bar h_{\mu\nu,\lambda} - 2\eta^{\mu\mu'}\eta^{\lambda\lambda'}
\bar{h}_{\mu_\lambda,\lambda'}\eta^{\sigma\sigma'}\\
&\bar{h}_{\mu'\sigma,\sigma'}{\big\}}
          + \frac{1}{2}{\big\{}\varphi_{,\mu}\varphi^{,\mu}-m_o^2\varphi^2 {\big\}} \\
&-\kappa {h}^{\mu\nu}{\big[}\overline{\varphi_{,\mu}\varphi_{,\nu}}+\frac{1}{2}m_o^2\varphi^2\eta_{\mu\nu}{\big{]}}\\
            & \quad - \kappa^2 [ \frac{1}{2}h_{\lambda\rho}\bar{h}^{\rho\lambda}{\big{(}} \varphi_{,\mu}\varphi^{,\mu} - m_o^2\varphi^2 {\big{)}} \\
&- 2\eta_{\rho\rho'}h^{\mu\rho}\bar{h}^{\rho'\nu}\varphi_{,\mu}\varphi_{,\nu}] + \cdots \\
\end{split}  
\label{eq1}
\end{equation}}\noindent
where $\varphi_{,\mu}\equiv \partial_\mu\varphi$ and 
we have
$g_{\mu\nu}(x)=\eta_{\mu\nu}+2\kappa h_{\mu\nu}(x)$,
$\eta_{\mu\nu}={\text diag}\{1,-1,-1,-1\}$ and 
$\bar y_{\mu\nu}\equiv \frac{1}{2}\left(y_{\mu\nu}+y_{\nu\mu}-\eta_{\mu\nu}{y_\rho}^\rho\right)$ for any tensor $y_{\mu\nu}$.
The Feynman rules for (\ref{eq1}) were 
already worked-out by Feynman~\cite{f1,f2},
where we use his gauge, $\partial^\mu \bar h_{\nu\mu}=0$. On this view,
quantum gravity is just another quantum field theory
where the metric now has quantum fluctuations as well.

For example, 
the one-loop corrections to the graviton propagator
due to matter loops is just given by the diagrams in Fig. 1.
\begin{figure}
\begin{center}
\epsfig{file=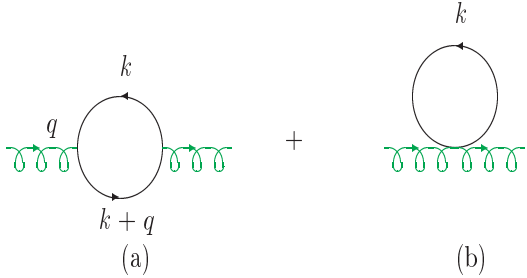,width=77mm,height=38mm}
\end{center}
\caption{\baselineskip=7mm     The scalar one-loop contribution to the
graviton propagator. $q$ is the 4-momentum of the graviton.}
\label{fig1}
\end{figure}
We return to these graphs shortly.

\section{Resummed Quantum Gravity}
In this section,
we will YFS resum the propagators in the theory:
from the YFS formula
\begin{equation}
iS'_F(p) = \frac{ie^{-\alpha B''_\gamma}}{S^{-1}_F(p)-{\Sigma'}_F(p)},
\label{yfsa}
\end{equation}
where ${\Sigma'}_F(p)$ is the sum of the YFS loop residuals,
we need to find for quantum gravity the analogue of
{\small
\begin{eqnarray}
\alpha B''_\gamma &= \int \frac{d^4\ell}{(2\pi)^4}\frac{-i\eta^{\mu\nu}}{(\ell^2-\lambda^2+i\epsilon)}\frac{-ie(2ik_\mu)}{(\ell^2-2\ell k+\Delta+i\epsilon)}\nonumber\\
&\frac{-ie(2ik'_\nu)}{(\ell^2-2\ell k'+\Delta'+i\epsilon)}{\Big|}_{k=k'},
\label{virt2}
\end{eqnarray}}
where $\Delta =k^2 - m^2$,~$\Delta' ={k'}^2 - m^2$ and
$\lambda$ is the IR cut-off. With the 
identifications~\cite{sw2} of the conserved 
graviton charges via 
$e \rightarrow \kappa k_\rho$ for soft emission from $k$
we get the analogue ,$-B''_g(k)$, of $\alpha B''_\gamma$
by replacing the $\gamma$ propagator in (\ref{virt2}) by the graviton 
propagator, 
and by replacing the QED charges by the corresponding gravity charges $\kappa k_{\bar\mu},~\kappa k'_{\bar\nu}$. 
This yields~\cite{bw1}
\begin{equation}
i\Delta'_F(k)|_{Resummed} =  \frac{ie^{B''_g(k)}}{(k^2-m^2-\Sigma'_s+i\epsilon)}.
\label{resum}
\end{equation}
with $B''_g(k) = \frac{\kappa^2|k^2|}{8\pi^2}\ln\left(\frac{m^2}{m^2+|k^2|}\right)$ in the deep Euclidean regime. If $m$ vanishes, using the usual $-\mu^2$ normalization 
point we get
$B''_g(k)=\frac{\kappa^2|k^2|}{8\pi^2}
\ln\left(\frac{\mu^2}{|k^2|}\right)$. In both cases the
resummed propagator falls faster than any power of $|k^2|$!
This is the basic result.
Note that
$\Sigma'_s$ starts in ${\cal O}(\kappa^2)$, so we may drop it in
calculating one-loop effects. 
This means that one-loop corrections are finite! 
Indeed, all quantum gravity loops are UV finite and the all orders
proof, as well as the explicit finiteness of ${\Sigma'}$
at one-loop, 
is given in Refs.~\cite{bw1}.

\section{ Newton's Law}

Consider the one-loop corrections to Newton's law
implied by the diagrams in Fig.~\ref{fig1}.
These corrections directly impact our black hole issue.
Introducing the YFS resummed propagators into Fig. 1 yields
, by the standard methods~\cite{bw1}, 
that the graviton propagator denominator,
$q^2 +\frac{1}{2}q^4\Sigma^{T(2)}+i\epsilon$, 
is specified by
$-\frac{1}{2}\Sigma^{T(2)} \cong \frac{c_2}{360\pi M_{Pl}^2}$
for
$c_2 = \int^{\infty}_{0}dx x^3(1+x)^{-4-\lambda_c x}\cong 72.1$ 
where $\lambda_c=\frac{2m^2}{\pi M_{Pl}^2}$.
This implies the potential
$\Phi_{N}(r)= -\frac{G_N M_1M_2}{r}(1-e^{-ar})$
where $a=1/\sqrt{-\frac{1}{2}\Sigma^{T(2)}}\simeq 3.96 M_{Pl}$
where for definiteness we set $m\cong 120$GeV.\par

We note for completeness that
$c_2 \cong \ln\frac{1}{\lambda_c}-\ln\ln\frac{1}{\lambda_c}-\frac{\ln\ln\frac{1}{\lambda_c}}{\ln\frac{1}{\lambda_c}-\ln\ln\frac{1}{\lambda_c}}-\frac{11}{6}$
and we used this result to check our numerical result for $c_2$.
Without resummation, $\lambda_c=0$,
our result for $c_2$ would be infinite.
Our gauge invariant result for $\Sigma^{T(2)}$
can be shown~\cite{bw1} to be consistent
with the one-loop analysis of QG in Ref.~\cite{thvelt1}.

Our deep Euclidean studies are complementary 
to the low energy studies
of Ref.~\cite{dono1}. The effective cut-off which we generate dynamically
is at $M_{Pl}$ so that renormalizable quantum field theory (QFT)  
below $M_{Pl}$
is unaffected. Some non-renormalizable QFT's are given new 
life here -- they may have other problems, however.

\section{Massive Elementary Particles and Black Holes}

Focusing the previous results, note that
,in the SM, there are 
now believed to be three massive neutrinos~\cite{neut},
with masses that we estimate at $\sim 3$ eV, and there are 
the remaining members
of the known three generations of Dirac fermions 
$\{e,\mu,\tau,u,d,s,c,b,t\}$. With reasonable estimates and measurements 
~\cite{pdg2002} of the SM particle masses, including the various bosons,
%
the result for $c_2$ for each
SM massive degree of freedom implies approximately
$c_{2,eff} \cong 9.26\times 10^3$
so that in the SM
$a_{eff} \cong 0.349 M_{Pl}$ .
To make direct contact with black hole physics, 
note that, if $r_S$ is the Schwarzschild radius,
for $r\rightarrow r_S$, $a_{eff}r \ll 1$ so 
that $|2\Phi_{N}(r)|_{m_1=m}/m_2|\ll 1$. This means that
$g_{00}\cong 1+2\Phi_{N}(r)|_{m_1=m}/m_2$ remains 
positive as we pass through the
Schwarzschild radius. 
It can be shown~\cite{bw1} that this 
positivity holds to $r=0$. Similarly, $g_{rr}$ remains negative
through $r_S$ down to $r=0$~\cite{bw1}. 
In resummed QG, a massive point particle is not a black hole.\par


Our results imply
$G_N(k)=G_N/(1+\frac{k^2}{a_{eff}^2})$
which is 
fixed point behavior for 
$k^2\rightarrow \infty$,
in agreement with the phenomenological asymptotic safety approach of
Ref.~\cite{reuter2}.
Our result that an elementary particle has no horizon
also agrees with the result in Ref.~\cite{reuter2} that a black hole
with a mass less than
 $M_{cr}\sim M_{Pl}$
has no horizon. The basic physics is the same: $G_N(k)$ vanishes for $k^2\rightarrow \infty$.

Because our value of the coefficient 
of $k^2$ in the denominator of $G_N(k)$
agrees with that found by Ref.~\cite{reuter2}, 
if we use their prescription for the
relationship between $k$ and $r$
in the regime where the lapse function
vanishes,
we get the same Hawking radiation phenomenology as
they do: a very massive black hole evaporates until it reaches a mass
$M_{cr}\sim M_{pl}$
at which the Bekenstein-Hawking temperature vanishes, 
leaving a Planck
scale remnant.


\section*{Acknowledgments}
We thank Prof. S. Jadach for useful discussions.
This work was partly supported by the US Department of Energy Contract  
DE-FG05-91ER40627
and by NATO Grants PST.CLG.977751,980342.


\begin{thebibliography}{99}

\bibitem{mtw} C. Misner, K.S. Thorne and J.A. Wheeler,
{\it Gravitation},( Freeman, San Francisco, 1973 ).
\bibitem{sw1} S. Weinberg, {\it Gravitation and Cosmology: Principles and Applications of the General Theory of Relativity},( John Wiley, New York, 1972).
\bibitem{abs} R. Adler, M. Bazin and M. Schiffer, {\it Introduction to General
Relativity },( McGraw-Hill, New York, 1965 ).
\bibitem{sm}S.L. Glashow, {\it Nucl. Phys.} {\bf 22} (1961) 579; 
S. Weinberg, {\it Phys. Rev. Lett.} {\bf 19} (1967) 1264;
A. Salam, in {\it Elementary Particle Theory}, ed. N. Svartholm
(Almqvist and Wiksells, Stockholm, 1968), p. 367;
G.~'t Hooft and M.~Veltman, {\it Nucl. Phys.}B {\bf 44}, 189 (1972)  
and {\bf B50}, 318 (1972); 
G.~'t Hooft, {\it ibid.} {\bf 35}, 167 (1971); M.~Veltman, {\it ibid.} {\bf 7}, 637 (1968);
D. J. Gross and F. Wilczek, 
{\it Phys. Rev. Lett.} {\bf 30} (1973) 1343;
H. David Politzer, {\it ibid.}{\bf 30} (1973) 1346; see also
, for example, F. Wilczek, in {\it Proc. 16th International Symposium on Lepton and Photon Interactions, Ithaca, 1993}, eds. P. Drell and D.L. Rubin 
(AIP, NY, 1994) p. 593, and references therein.
\bibitem{gsw} See, for example, M. Green, J. Schwarz and E. Witten,
{\it Superstring Theory, v. 1 and v.2}, ( Cambridge Univ. 
Press, Cambridge, 1987 ) and references therein.
\bibitem{jp}
See, for example, J. Polchinski, {\it String Theory, v. 1 and v. 2},
(Cambridge Univ. Press, Cambridge, 1998), and references therein.
\bibitem{lqg}
See for example V.N. Melnikov, {\it Gravit. Cosmol.} {\bf 9}, 118 (2003);
L. Smolin, hep-th/0303185, and references therein.
\bibitem{bw1} B.F.L. Ward,{\it Mod. Phys. Lett.}A{\bf 17}, 2371 (2002); 
{\it ibid.}{\bf 19} (2004)143; {\it J. Cos. Astropart. Phys.}{\bf 0402}, 011 (2004).
\bibitem{f1} R. P. Feynman, {\it Acta Phys. Pol.} {\bf 24}, 697 (1963).
\bibitem{f2} R. P. Feynman, {\it Feynman Lectures on Gravitation},
eds. F.B. Moringo and W.G. Wagner (Caltech, Pasadena, 1971).
\bibitem{wein1}
S. Weinberg, in{\it General Relativity}, eds. S.W. Hawking
and W. Israel,( Cambridge Univ. Press, Cambridge, 1979) p.790.
\bibitem{laut} O. Lauscher and M. Reuter, hep-th/0205062, and references
therein.
\bibitem{reuter2}
A. Bonnanno and M. Reuter, {\it Phys. Rev.} D{\bf 62}, 043008 (2000).
\bibitem{yfs}D.~R.~Yennie, S.~C.~Frautschi, and H.~Suura, Ann. Phys. {\bf 13}, 379 (1961);
see also K.~T.~Mahanthappa, {\it Phys.~Rev.}~{\bf 126}, 329 (1962), for a related analysis.
\bibitem{yfs1} See S. Jadach {\it et al.},{\it Comput. Phys. Commun.} {\bf 102},229 (1997); S. Jadach, M. Skrzypek and B.F.L. Ward,{\it Phys.Rev.} D{\bf 55},1206 (1997); S. Jadach, B.F.L. Ward and Z. Was,{\it Phys. Rev.} D{\bf 63}, 113009 (2001);
S. Jadach, B.F.L. Ward and Z. Was, {\it Comp. Phys. Commun.} {\bf 130}, 260 (2000);
S. Jadach {\it et al.}, {\it ibid.}{\bf 140} 432, 475, (2001).
\bibitem{hawk} S. Hawking, Nature ( London ) {\bf 248}, 30 (1974);
Commun. Math. Phys. {\bf 43}, 199 ( 1975 ).
\bibitem{mlg} M.L. Goldberger, private communication, 1972.
\bibitem{lewwg}
D. Abbaneo {\it et al.}, hep-ex/0212036; see also, M. Gruenewald, 
hep-ex/0210003, in {\it Proc. ICHEP02}, eds. S. Bentvelsen {\it et al.} ( North-Holland, Amsterdam, 2003) 280.
\bibitem{sw2}
S. Weinberg, {\it The Quantum Theory of Fields, vols. 1-3}, 
( Cambridge University Press, Cambridge, 1995,1996,2000).
\bibitem{thvelt1}
G. 't Hooft and M. Veltman, {\it Ann. Inst. Henri Poincare} 
{\bf XX}, 69 (1974).
\bibitem{dono1} See for example J. Donoghue, 
{\it Phys. Rev. Lett.} {\bf 72}, 2996 (1994);
{\it Phys. Rev.} D{\bf 50}, 3874 (1994); preprint gr-qc/9512024; 
J. Donoghue {\it et al.}, {\it Phys. Lett.} B{\bf 529}, 132 (2002), 
and references therein.
\bibitem{neut} See for example D. Wark, in {\it Proc. ICHEP02}, eds. S. Bentvelsen {\it et al.} (North-Holland, Amsterdam, 2003) 164; 
M. C. Gonzalez-Garcia, {\it op. cit.} 186, hep-ph/0211054,
and references therein.
\bibitem{pdg2002}K. Hagiwara {\it et al.}, {\it Phys. Rev.} D{\bf 66}, 010001 (2002)1; see also H. Leutwyler and J. Gasser, {\it Phys. Rept.} {\bf 87}, 77 (1982), and references therein.
\end{thebibliography}
\end{document}